\documentclass[a4paper,12pt]{article}
\usepackage{amssymb, amsmath}
\usepackage[dvips]{graphicx, color}

\usepackage{textcomp} 

\newcommand{\denomi}[1]{$\lozenge#1$}
\newcommand{\mdenomi}[1]{\lozenge#1} 

\usepackage{cite}

\title{Fractal behind smart shopping}
\author{Ken Yamamoto and Yoshihiro Yamazaki}
\date{\it Waseda University, 169-8555, Japan}

\begin{document}
\maketitle

\begin{abstract}
The `minimal' payment---a payment method
which minimizes the number of coins in a purse---is presented.
We focus on a time series of change given back to a shopper
repeating the minimal payment.
The delay plot shows visually that
the set of successive change possesses
a fine structure similar to the Sierpinski gasket.
We also estimate effectivity of the minimal-payment method
by means of the average number of coins in a purse,
and conclude that the minimal-payment strategy
is the best to reduce the number of coins in a purse.
Moreover, we compare our results to
the rule-60 cellular automaton and the Pascal-Sierpinski gaskets,
which are known as generators of the discrete Sierpinski gasket.
\end{abstract}

\section{Introduction}
An English proverb says that ``a heavy purse makes a light heart'',
but we do not welcome a literal heavy purse full of coins.
By repeated payments in everyday shopping,
the number of coins in the shopper's purse constantly increases or decreases,
and sometimes the purse gets full of coins.
A purse with a lot of coins is uneconomic
because it becomes heavy and bulky, and even worse,
the shopper has a lot of trouble in picking out coins from the purse.
Electronic money\cite{Furche} is hoped to eliminate such trouble,
but it has not become a substitute for real cash so far
because of technical, economical, and legal issues.

Some mathematical problems treat of coins.
A famous one is known as the change-making problem\cite{Chang}:
how can a given amount of money be made
with the least number of coins of given denominations?
This problem is a variation of the knapsack problem\cite{Martello},
and it is known to be NP-hard\cite{Hromkovic}.
The greedy algorithm\cite{Magazine} gives a solution in some
limited cases,
and other heuristic algorithms have been
also proposed\cite{Wright, Pearson, Bocker}.
The optimality or efficiency of a monetary system itself
has been also evaluated\cite{Telser, Tschoegl, Shallit},
and the Frobenius coin problem\cite{Nijenhuis, Kannan, Beck}
asks 
what is the largest amount of money
that cannot be obtained using only coins of specified denominations.
We note that such problems are for retailers (how to make change)
or for governments (what currency system is preferable);
consumers are not considered.
To prevent the swelling of a purse, which is a problem for consumers,
we propose a method to minimize the number of coins in the purse
through the payment.
We also show that this method accompanies a remarkable mathematical structure
like fractals.

\section{What is the minimal payment?}
We define the `minimal payment' as follows.
(i) A retailer keeps all kinds of coins sufficiently,
and gives back change with the least number of coins.
Namely, the retailer never gives 10-cent change
by ten 1-cent coins or two 5-cent coins.
(ii) The shopper pays with a combination of coins
that minimizes the number of coins in the purse after the payment.
(iii) The coins paid by the shopper and the ones given back to the shopper
do not include common coins.

\begin{figure}[p!] \centering
\includegraphics[width=89mm]{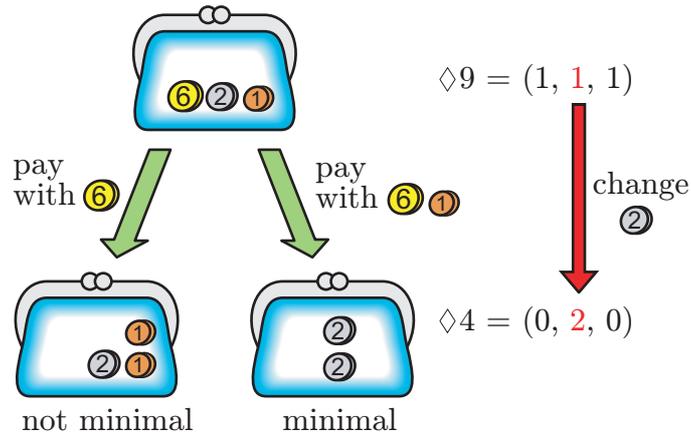}
\caption{
Illustration for the minimal payment in a simple situation:
coins are \denomi{6}, \denomi{2}, and \denomi{1}.
The notation $(1,1,1)$ means that
one \denomi{6}-, one \denomi{2}-, and one \denomi{1}-coins
are present in the purse.
A shopper is buying an item of \denomi{5}.
Compare two types of payment (left: not minimal, right: minimal).
}
\label{fig:purse}
\end{figure}

Let us play with a simple example.
Imagine a nation where the coins are issued in
\denomi{6}, \denomi{2}, and \denomi{1} denominations
(`\denomi{}' is the currency sign of this nation, and
we also use this symbol as the generic currency sign in this paper,
instead of the formal symbol `\textcurrency'),
and a shopper in this nation has each of \denomi{6}, \denomi{2},
and \denomi{1} coins in the purse,
and is buying an item of \denomi{5}.
We compare two ways of payment.
(1) If the shopper pays with one \denomi{6}-coin alone,
then the change is \denomi{1},
hence three coins (one \denomi{2}- and two \denomi{1}-coins)
are left in the purse.
(2) If the shopper pays with one \denomi{6}- and one \denomi{1}-coins,
then the change is \denomi{2} (paid with {\bf one} \denomi{2}-coin),
hence two \denomi{2}-coins are left after the payment.
Other payment patterns do not meet the condition (iii).
Therefore,
the minimal payment in this case is that
the shopper pays with one \denomi{6}- and one \denomi{1}-coins
and receives one \denomi{2}-coin in change
(see Fig. \ref{fig:purse}).
The important fact is that
the number of coins left in the shopper's purse
can vary according to the combination of coins in the payment.

Next we present the efficient calculation method of the minimal payment.
Let us consider again the currency `\denomi{}'.
The notation $(n_1, n_2, n_3)$ represents
the content of the shopper's purse,
where $n_1$, $n_2$, and $n_3$ are respectively
the number of \denomi{6}, \denomi{2}, and \denomi{1} coins.
In the above example, the coins before the payment are
written as $(1, 1, 1)$ (see also Fig. \ref{fig:purse}).
By using this vector-like notation,
the total amount of money in the purse can be expressed
as the scalar product:
$(1,1,1)\cdot(\mdenomi{6}, \mdenomi{2},\mdenomi{1})
=1\cdot\mdenomi{6}+1\cdot\mdenomi{2}+1\cdot\mdenomi{1}=\mdenomi{9}$.
In this sense, we identify a vector with the total amount,
and simply express as $\mdenomi{9}=(1,1,1)$.
Returning to the above example,
\denomi{4}($=\mdenomi{9}-\mdenomi{5}$) is left in the purse
after the purchase of an item of \denomi{5}.
Through the
transition from $\mdenomi{9}=(1,1,1)$ (before the payment)
to $\mdenomi{4}=(0,2,0)$ (after the payment),
one \denomi{6}- and one \denomi{1}-coins are gone
and one \denomi{2}-coin is gained,
which correspons to the minimal payment process (Fig. \ref{fig:purse}).
We note that
this calculation procedure for the minimal payment
is far more efficient than the `full search',
and it can be applied straightforwardly to general currencies.

\section{Fractality of successive minimal-payment processes}
\label{sec:3}
Here we show the mathematical structure of
successive minimal-payment processes.
We set the following four assumptions.
(i) The shopper and the retailer iterate the minimal payment.
(ii) Prices are generated randomly and independently in each payment,
and we used a uniform random number in the numerical calculations.
(iii) In order for the shopper not to lack money,
we assume that the shopper has sufficiently
large number of banknotes, and that
the shopper can use banknotes if needed.
(iv) Every coin is a multiple of the immediately smaller coin,
and the smallest banknote is a multiple of the largest coin.
We call (iv) the `multiplicable condition.'
For example,
the Japanese yen
(1000 yen bill; 500, 100, 50, 10, 5, 1 yen coins),
the Korean won
(1000 won bill; 500, 100, 50, 10, 5, 1 won coins),
and the Swedish krona
(20 krona bill; 10, 5, 1 krona, 50 \"ore (= 0.5 krona) coins)
are multiplicable.
However,
the majority of currencies in the world
is partially not multiplicable;
in the US dollar
(1 dollar (= 100 cent) bill; 50, 25, 10, 5, and 1 cent coins),
25 is not a multiple of 10.

For the analysis of successive change in the minimal payment,
we employ the delay plot.
For a given time series $z(0), z(1),\cdots,z(T)$,
the delay plot is defined as a planar point set
$\{(z(t),z(t+1))\vert t=0,1,\cdots, T-1\}$.
The delay plot is a useful tool for
visualizing the correlation of adjoining pairs of the time series.
In chaos theory\cite{Alligood} and
nonlinear time series analysis\cite{Kantz},
the delay plot is used for the reconstruction of
an attractor from a series of experimental data.
In general,
a delay plot generated from a deterministic system
forms an organized structure,
whereas a plot from a stochastic noisy system
forms an indistinct distribution.

\begin{figure}[p!] \centering
\includegraphics[width=89mm]{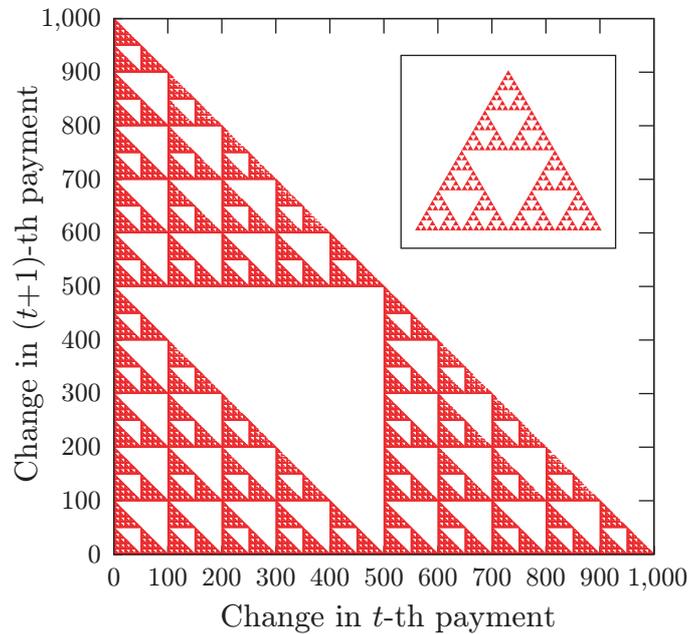}
\caption{
The delay plot of a series of successive change in the Japanese yen
(1000 yen bill; 500, 100, 50, 10, 5, 1 yen coins),
constructed from a minimal-payment simulation iterated $10^7+1$ times.
Inset: the Sierpinski gasket.
}
\label{fig:gasket-yen}
\end{figure}

Figure \ref{fig:gasket-yen} illustrates the delay plot
of a series of change
generated by computer simulation of the minimal payment.
In the simulation,
we adopted the Japanese yen as the currency system
(which holds the multiplicable condition),
and iterated payment processes $10^7+1$ times.
The whole structure is similar to the celebrated fractal object
`Sierpinski gasket'\cite{Sierpinski}.

In order to explain the mechanism of this quasi-fractal structure,
let us go back once again to the model currency `\denomi{}',
where \denomi{6}, \denomi{2}, and \denomi{1} coins are issued.
We additionally imagine that the smallest banknote is
\denomi{12} denomination.
Note that this currency is multiplicable.
$(n_1(t), n_2(t), n_3(t))$ represents the coins in the shopper's purse
after the $t$-th payment,
and we set the initial condition $(n_1(0), n_2(0), n_3(0))=(0,0,0)$.
We also denote by $c(t)=(c_1(t), c_2(t), c_3(t))$
the change given back in the $t$-th payment.
Since the prices are randomly generated,
$(n_1,n_2,n_3)$ and $(c_1,c_2,c_3)$ are random variables.
If $n_1(t)>n_1(t-1)$ (\denomi{6} coins in the purse increases),
the $t$-th change includes $n_1(t)-n_1(t-1)$ pieces of \denomi{6} coin.
Otherwise, if $n_1(t)\le n_1(t-1)$,
the $t$-th change includes no \denomi{6} coins.
Hence
we can simply express as
$c_1(t)=\max\{n_1(t)-n_1(t-1), 0\}$.
The sum of the number of \denomi{6} coins in $t$-th and $(t+1)$-th change
is then calculated as
\begin{align}
c_1(t)+c_1(t+1)
&=\max\{n_1(t)-n_1(t-1),0\}+\max\{n_1(t+1)-n_1(t),0\} \nonumber\\
&=\max\{n_1(t+1)-n_1(t-1), n_1(t)-n_1(t-1), n_1(t+1)-n_1(t), 0\},
\label{eq:1}
\end{align}
where we use the formula in a max-plus algebra\cite{Baccelli}:
$\max\{a, b\}+\max\{c, d\}=\max\{a+c, a+d, b+c, b+d\}$.
By the definition of the minimal payment and above algorithm,
we can derive
$0\le n_1(t)\le 1$ for any $t\ge0$,
where `1' at the right hand side
is given by the ratio of two denominations as
$\mdenomi{12}/\mdenomi{6}-1$.
Therefore, the four arguments
of `$\max$' in Eq. \eqref{eq:1}
are all equal to or less than $1$ constantly,
and we obtain $0\le c_1(t)+c_1(t+1)\le 1$ as a result.
Similarly, we can derive the relations
$0\le c_2(t)+c_2(t+1)\le 2 (=\mdenomi{6}/\mdenomi{2}-1)$ and
$0\le c_3(t)+c_3(t+1)\le 1 (=\mdenomi{2}/\mdenomi{1}-1)$
for $c_2$ and $c_3$ respectively.
These relations signify that
{\bf successive change $c(t)$ and $c(t+1)$ are correlated}.

\begin{figure}[p!] \centering
\hbox to \hsize{\hss
\includegraphics[width=180mm]{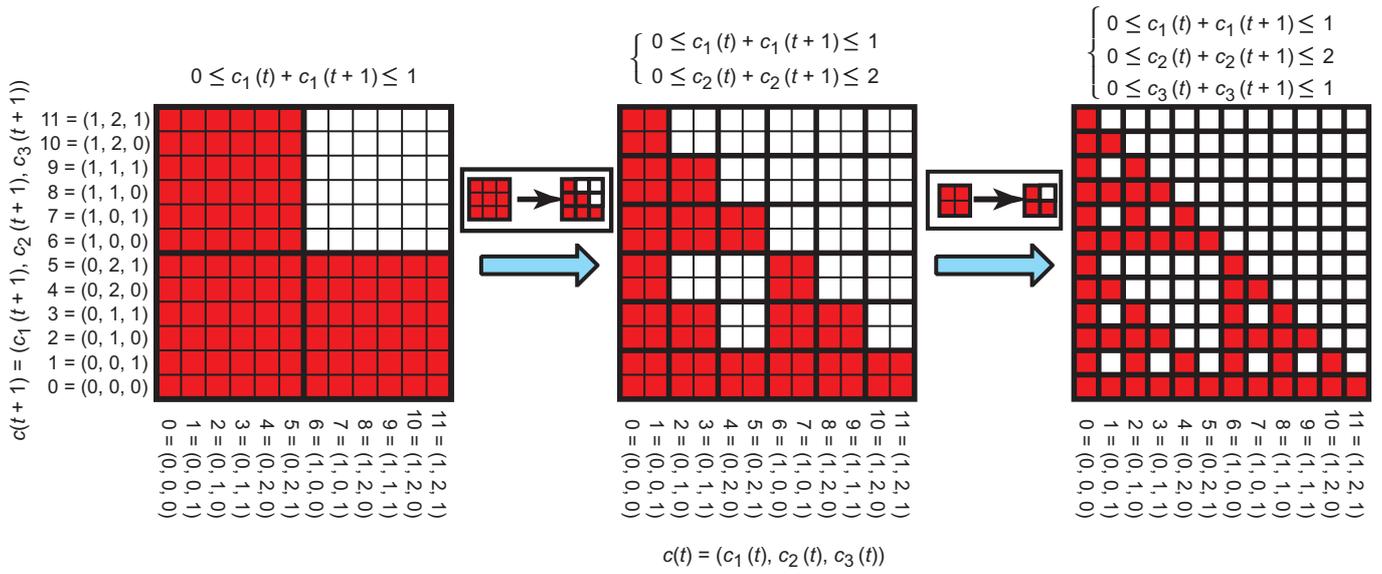}\hss}
\caption{
The iterative steps for the determination of the possible area
of the delay plot in the case of
the currency `\denomi{}'
(\denomi{12} bill; \denomi{6}, \denomi{2}, and \denomi{1} coins).
The pattern gets higher resolution
as conditions for smaller coins are included.
The unit squares are used to indicate lattice points.
}
\label{fig:iteration}
\end{figure}

\begin{figure}[p!] \centering
\includegraphics[width=89mm]{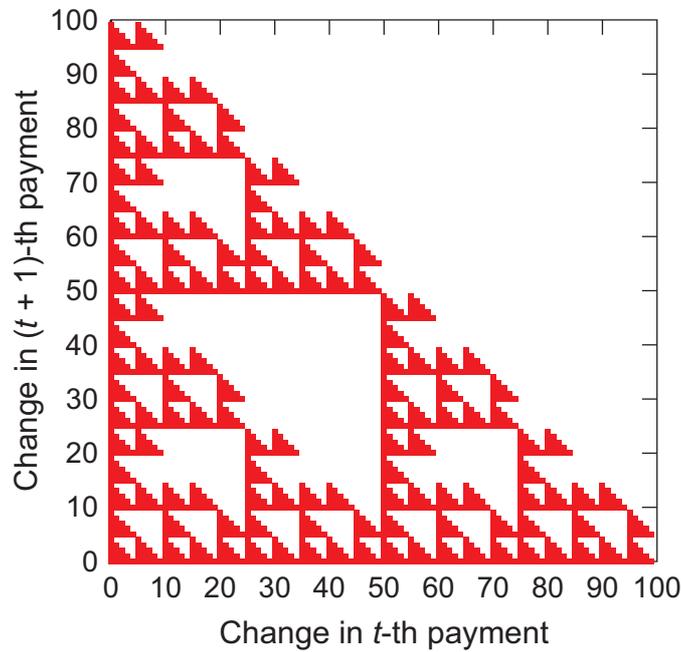}
\caption{
The delay plot of a non-multiplicable case:
the US dollar
(1 dollar bill; 50, 25, 10, 5, and 1 cent coins).
}
\label{fig:gasket-cent}
\end{figure}

By the above calculation,
the points $(c(t), c(t+1))$ of the delay plot
are confined within the lattice points
satisfying the conditions
\begin{equation}
\left\{
\begin{array}{l}
0\le c_1(t)+c_1(t+1)\le1,\\
0\le c_2(t)+c_2(t+1)\le2,\\
0\le c_3(t)+c_3(t+1)\le1.
\end{array}
\right.
\label{eq:area}
\end{equation}
As in Fig. \ref{fig:iteration},
the effect of including smaller coins progressively
generates a nested pattern
(smaller coins contribute to higher resolution).
In each iterative step,
the blocks (the squares with bold contours)
are replaced with smaller blocks aligned in a step-shape (or an L-shape),
and finally the structure of the delay plot has a fine structure
as in the Sierpinski gasket.
Note that the side lengths, 12 (the length of the `whole' block),
6, 2, and 1, of blocks in the iterative steps
are identical with the denominations
\denomi{12}, \denomi{6}, \denomi{2}, and \denomi{1}.
This scenario of `successive replacement' universally works
on general multiplicable currency systems.
In particular, if the denominations of coins grow geometrically
(i.e., \denomi{1}, \denomi{r}, \denomi{r^2}, $\cdots$),
the delay plot possesses (discrete) self-similarity,
and its fractal dimension $D$ is given by
\begin{equation}
D=\frac{\ln\frac{r(r+1)}{2}}{\ln r}=1+\frac{\ln\frac{r+1}{2}}{\ln r}.
\label{eq:dimension}
\end{equation}
This expression also appears in a study of Pascal's triangle\cite{Gamelin};
we discuss delay plots and Pascal's triangle in more detail in Section \ref{sec:Pascal}.
The multiplicable condition is necessary for the
realization of a regular shape of the delay plot.
For example,
the delay plot of the US dollar
exhibits a relatively disordered structure,
reflecting the breaking of the multiplicable condition
between 25 and 10 cent coins
(Fig. \ref{fig:gasket-cent}).

\section{Effectivity of the minimal payment}
\begin{figure}[p!] \centering
\includegraphics[width=170mm]{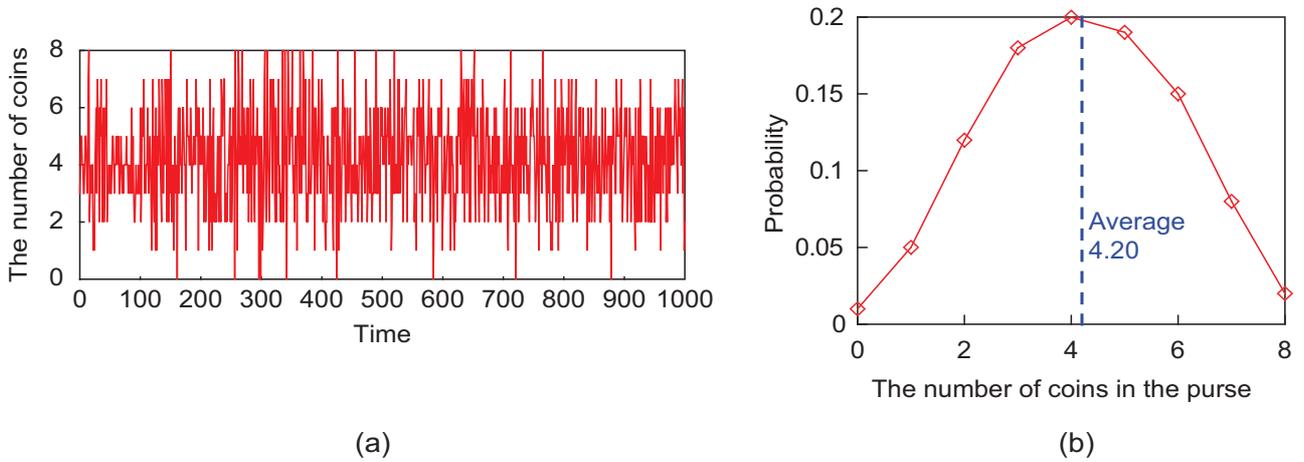}
\caption{
(a) The time siries of the number of coins in the shopper's purse,
generated by a numerical simulation of the minimal payment.
The currency is the US dollar (1 dollar bill; 50, 25, 10, 5, and 1 cent coins),
and the first 1000 steps are shown.
(b) The numerically-computed probability distribution
of the number of coins in the purse.
The average of this distribution is 4.20 pieces, indicated by the dashed line.
}
\label{fig:timeseries}
\end{figure}

Here we comment on the effectivity of the minimal-payment method,
measured by the average number of coins in the shopper's purse.
We estimated this average numerically.
In this calculation, we adopted the US dollar as the currency,
and iterated the minimal payment $10^7$ times.
The results are summarized in Fig. \ref{fig:timeseries}.
Figure \ref{fig:timeseries} (a) shows the fluctuation
of the number of coins in the purse of the first 1000 steps,
and Fig. \ref{fig:timeseries} (b)
is the numerically-generated probability distribution
of the number of coins in the purse.
By taking the mean value of Fig. \ref{fig:timeseries} (b),
the number of coins in the purse is 4.20 pieces on average.
Furthermore,
the maximum number of coins in Fig. \ref{fig:timeseries} (b) is eight.
Thus, by using the minimal-payment strategy,
we can constantly keep the coins in a purse 8 pieces or less.

\begin{figure}[p!] \centering
\includegraphics[width=160mm]{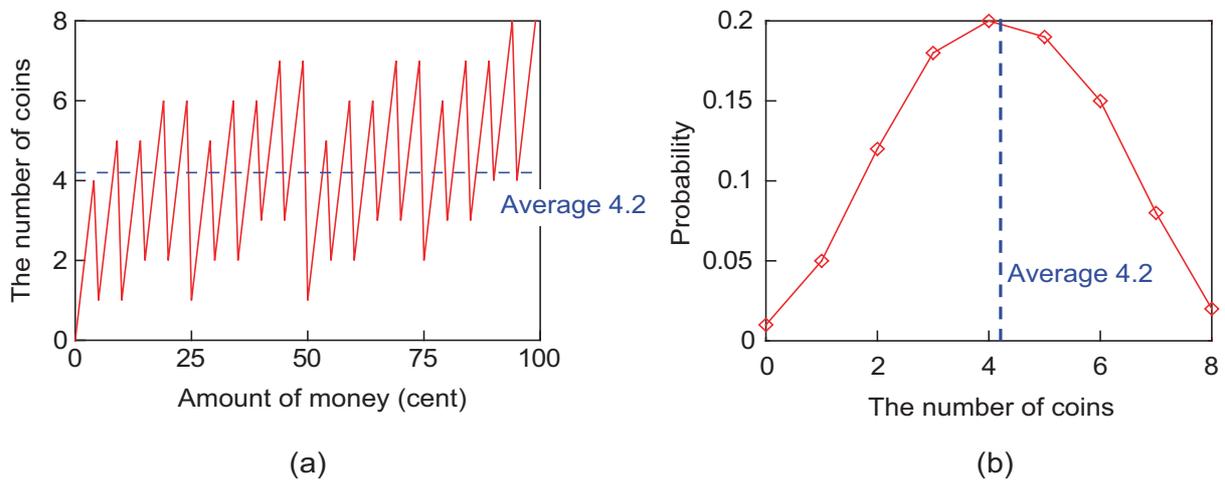}
\caption{
(a) The least number of coins that
the US dollar needs to represent each amount of money.
(b) The distribution of the number of coins, built from (a).
The average number of coins is 4.2 pieces,
indicated by the dashed lines.
}
\label{fig:rigorous_dollar}
\end{figure}

We can also calculate this average more rigorously.
Since the prices are distributed randomly and independently
in our numerical model,
a probability of the amount of money in the purse
becomes uniform.
Figure \ref{fig:rigorous_dollar} illustrates
how many coins are needed to represent each amount of money,
and the average number of coins is exactly 4.2 pieces by taking the mean value.
Note that the two distributions in
Fig. \ref{fig:timeseries} (b) and Fig. \ref{fig:rigorous_dollar} (b)
are so alike.

The average number can be calculated easily for a multiplicable currency.
For the model currency `\denomi{}'
(\denomi{12} bill; \denomi{6}, \denomi{2}, and \denomi{1} coins),
the set of possible amounts of money in a purse is
\begin{equation}
\begin{array}{llll}
\mdenomi{0}=(0,0,0), & \mdenomi{1}=(0,0,1),
	& \mdenomi{2}=(0,1,0), & \mdenomi{3}=(0,1,1),\\
\mdenomi{4}=(0,2,0), & \mdenomi{5}=(0,2,1),
	& \mdenomi{6}=(1,0,0), & \mdenomi{7}=(1,0,1),\\
\mdenomi{8}=(1,1,0), & \mdenomi{9}=(1,1,1),
	& \mdenomi{10}=(1,2,0), & \mdenomi{11}=(1,2,1).
\end{array}
\label{eq:list}
\end{equation}
For the same reason as above,
each amount of money listed in Eq. \eqref{eq:list}
emerges with a uniform probability.
In Eq. \eqref{eq:list},
the possible numbers of \denomi{6} coins are either 0 or 1,
which appear six times each.
Thus, the average number of \denomi{6} coins is 0.5,
which is also given by $(\mdenomi{12}/\mdenomi{6}-1)/2$.
The same estimation is valid for
the numbers of \denomi{2} and \denomi{1} coins.
Therefore, the average number of coins in a purse is
\[
\frac{\mdenomi{12}/\mdenomi{6}-1}{2}+\frac{\mdenomi{6}/\mdenomi{2}-1}{2}
+\frac{\mdenomi{2}/\mdenomi{1}-1}{2}=2.
\]
We can straightforwardly apply this formula
to any multiplicable currencies.
For example,
the average number of the Japanese yen
(1000 yen bill; 500, 100, 50, 10, 5, and 1 yen coins) is both 7.5 pieces.

Conversely,
we note that this average number can be regarded as the lower bound,
that is,
we cannot keep down the average number of coins below this lower bound,
no matter what payment strategy is used.
This is the `minimality' of the presented payment method.

\section{Relevance to the rule-60 cellular automaton}
\begin{figure}[p!] \centering
\includegraphics[width=80mm]{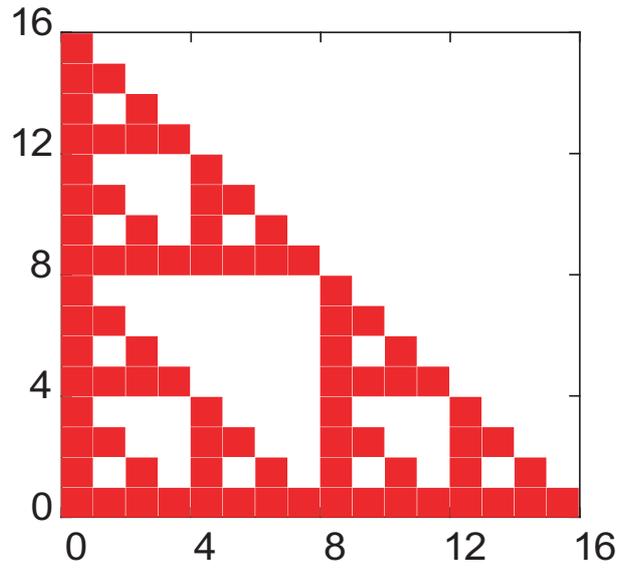}
\caption{
A delay plot of a binary currency
(the coins are \denomi{1}, \denomi{2}, \denomi{4}, \denomi{8};
and the smallest banknote is \denomi{16})
forms a discrete Sierpinski gasket.
}
\label{fig:gasket-binary}
\end{figure}

In the present section,
let us consider the denominations of coins grow with a binary scale,
that is, the coins are \denomi{1}, \denomi{2}, \denomi{4}, \denomi{8},
$\cdots$, \denomi{2^{n-1}}, and the smallest banknote is \denomi{2^n}.
As in Section \ref{sec:3},
let $c_i(t)$ denote the number of \denomi{2^{n-i}} coins in $t$-th change,
and the condition $c_i(t)=0 \mbox{ or }1$ holds for any $t$ and $i$.
The corresponding delay plot is obtained by solving
\begin{equation}
\left\{
\begin{array}{c}
0\le c_1(t)+c_1(t+1)\le 1, \\
0\le c_2(t)+c_2(t+1)\le 1, \\
\vdots\\
0\le c_n(t)+c_n(t+1)\le 1.
\end{array}
\right.
\label{eq:area-binary}
\end{equation}
In other words, $c_n(t)=1$ and $c_n(t+1)=1$ never take place at the same time.
The delay plot forms the discrete Sierpinski gasket
(Fig. \ref{fig:gasket-binary}).

The discrete Sierpinski gasket appears in various systems
in mathematics and physics,
ranging from an analysis of the Tower of Hanoi\cite{Hinz}
to pulse dynamics in a reaction-diffusion system\cite{Hayase} and a model of fracture\cite{Matsushita}.
As a typical example,
we focus on the rule-60 cellular automaton in this section.

\begin{figure}[p!] \centering
\includegraphics[width=120mm]{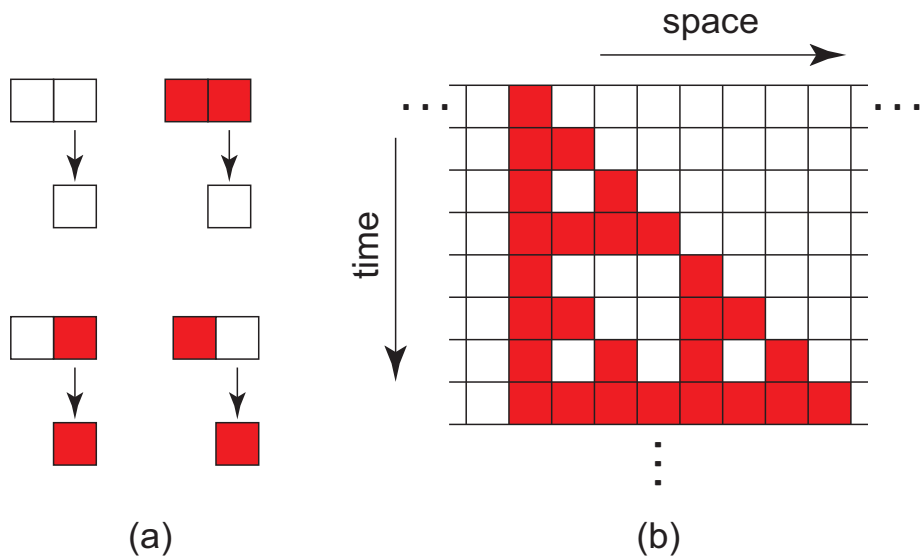}
\caption{
(a) Evolution of the rule-60 cellular automaton.
White cells represent the state `0', and colored ones represent the state `1'.
(b) A spatio-temporal pattern of the rule-60 cellular automaton.
Horizontal direction represents space,
and vertical direction represents time.
Initially,
one cell is the state 1, and the others are 0.
}
\label{fig:rule60}
\end{figure}

A cellular automaton (CA)\cite{Wolfram} is
a discrete dynamical system on cells aligned linearly.
The possible state of each cell is either `0' or `1',
and the evolution of states generally depends on their own state
and the states of their neighbors.
In particular, the rule-60 CA can generate the discrete Sierpinski gasket,
which is driven by the following rule.
(i) If a cell and its left neighbor are in the same state,
the next state of the cell is 0.
(ii) Otherwise (i.e., the states of the two cells are different),
the next state is 1.
This evolution rule is schematically shown in Fig. \ref{fig:rule60} (a).
We denote $s_n^t$ by the state of $n$-th cell at time $t$.
Then, the evolution of the state is simply expressed as
$s_n^{t+1}=s_n^t\oplus s_{n-1}^t$,
where $\oplus$ is an exclusive-or operation
defined as $0\oplus0=1\oplus1=0$ and $0\oplus1=1\oplus0=1$.
A resultant pattern is depicted in Fig. \ref{fig:rule60} (b),
where the initial condition is that
one cell is `1' and the others are `0'.
The evolution rules of other cellular automata,
such as rule-90, 22, 154 and 102,
can be expressed by $\oplus$,
and they also generate discrete Sierpinski gaskets.

\begin{figure}[p!] \centering
\includegraphics[width=150mm]{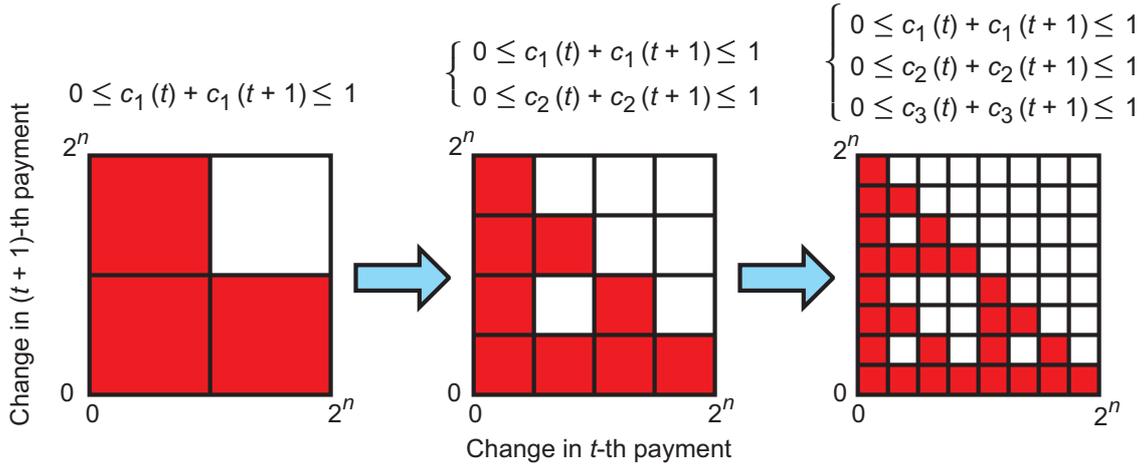}
\caption{
The first three iterative steps for solving Eq. \eqref{eq:area-binary}.
We can reach the solution
by simply applying the rule \eqref{eq:rule} to each block recursively.
}
\label{fig:iteration-binary}
\end{figure}

\begin{figure}[p!] \centering
\includegraphics[width=140mm]{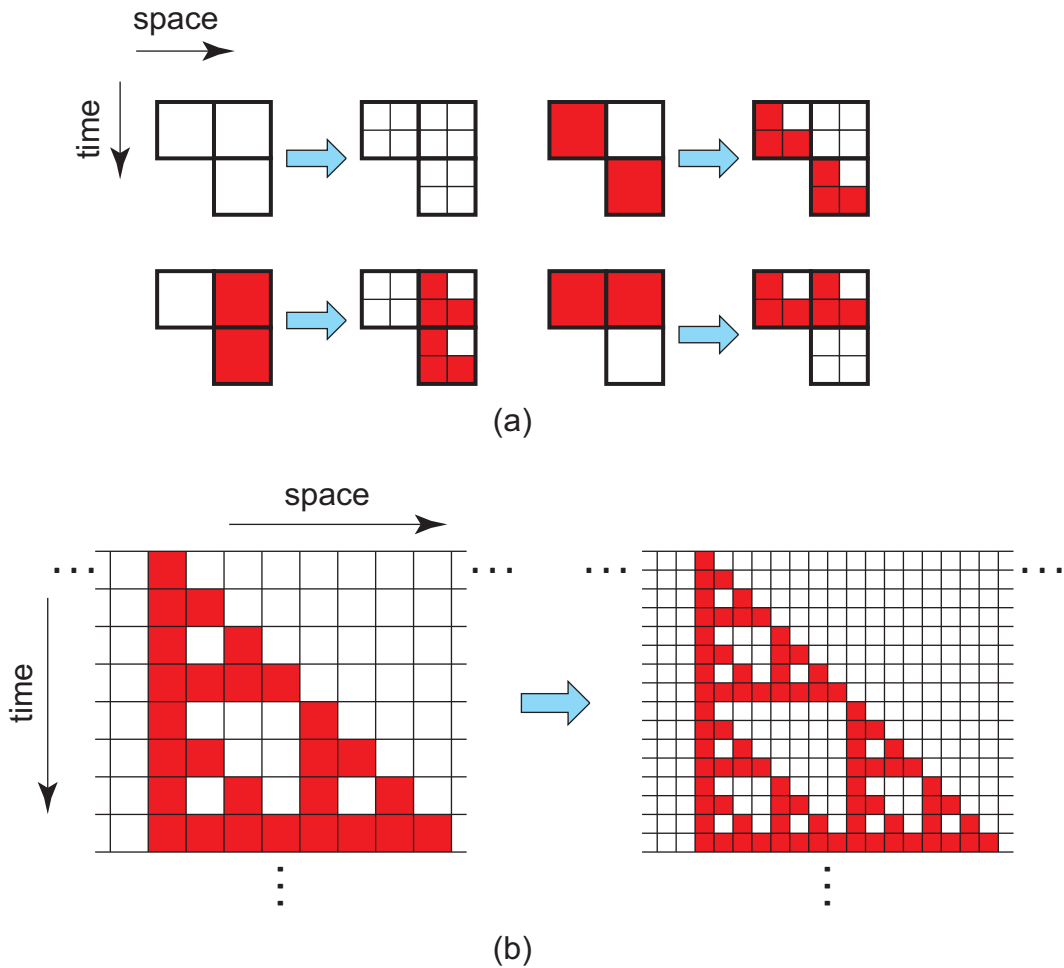}
\caption{
(a) Subdivision of the elementary evolution rule
of the rule-60 cellular automaton.
The small cells again possess the exclusive-or structure.
(b) Subdivision of a spatio-temporal pattern of Fig. \ref{fig:rule60} (b),
acting as a scale transformation.
}
\label{fig:subdivision}
\end{figure}

Similarity between a delay plot of binary coins (Fig. \ref{fig:gasket-binary})
and the evolution pattern of the rule-60 CA
(Fig. \ref{fig:rule60} (b)) is obvious.
We discuss this similarity more in detail.
Figure \ref{fig:iteration-binary} shows iterative steps
for solving Eq. \eqref{eq:area-binary}.
As a condition for a smaller coin is considered,
the blocks in each step are subdivided according to the following rule:
\newlength{\depthsubdivision}
\begin{equation}
\setlength{\depthsubdivision}{-0.25\baselineskip}
\raisebox{\depthsubdivision}{
\includegraphics[height=\baselineskip]{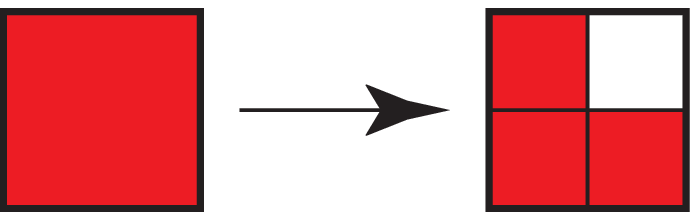}
}
\quad\mbox{and}\quad
\raisebox{\depthsubdivision}{
\includegraphics[height=\baselineskip]{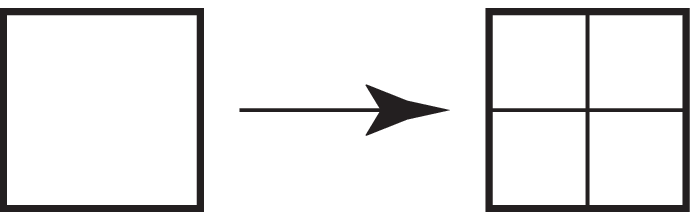}
}.
\label{eq:rule}
\end{equation}
As shown in Fig. \ref{fig:subdivision} (a),
this subdivision preserves the exclusive-or structure
of the rule-60 CA.
Moreover, a spatio-temporal pattern after the subdivision
is similar to the one before the subdivision,
that is, the subdivision works on the rule-60 CA
as a spatio-temporal scale transformation.
Both the unit lengths of space and time are halved by subdivision
(see Fig. \ref{fig:subdivision} (b)).
Therefore,
we conclude that the key to similarity between
the delay plot of a binary currency and the rule-60 CA
is
compatibility of the rule \eqref{eq:rule} of subdivision
and the exclusive-or relation.

\section{Comparison with the Pascal-Sierpinski gaskets}\label{sec:Pascal}
\begin{figure}[p!] \centering
\includegraphics[width=100mm]{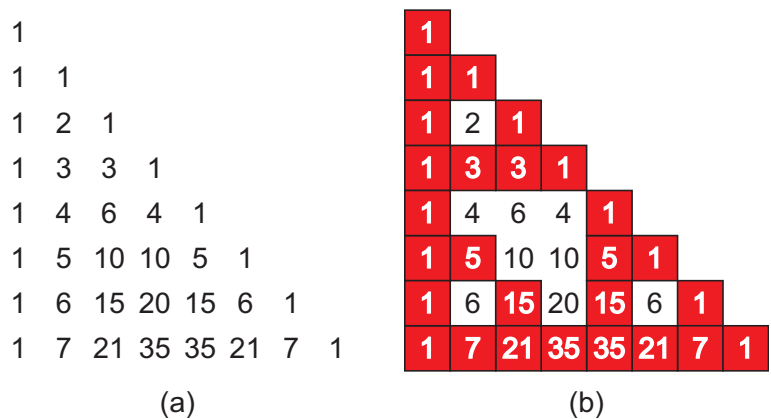}
\caption{
(a) A part of Pascal's triangle.
The elements are aligned on the square lattice.
(b) The discrete Sierpinski gasket is obtained by coloring the odd numbers.
}
\label{fig:pascal}
\end{figure}

Pascal's triangle (Fig. \ref{fig:pascal} (a))
also generates a discrete Sierpinski gasket
by coloring the odd numbers (Fig. \ref{fig:pascal} (b))\cite{Wolfram1984}.
Clearly, Fig. \ref{fig:pascal} (b) is essentially the same
as Fig. \ref{fig:gasket-binary} and Fig. \ref{fig:rule60} (b).
In fact, the coloring rule is reduced
to the exclusive-or operation\cite{Culik}.
As a generalization,
the Pascal-Sierpinski gasket\cite{Holter} of modulo $r$ is obtained
by coloring the elements not divisible by $r$.
In this section, we compare the Pascal-Sierpinski gasket of modulo $r$
and the delay plot of the coins
\denomi{1}, \denomi{r}, \denomi{r^2}, $\cdots$.

\begin{figure}[p!] \centering
\includegraphics[width=.9\textwidth]{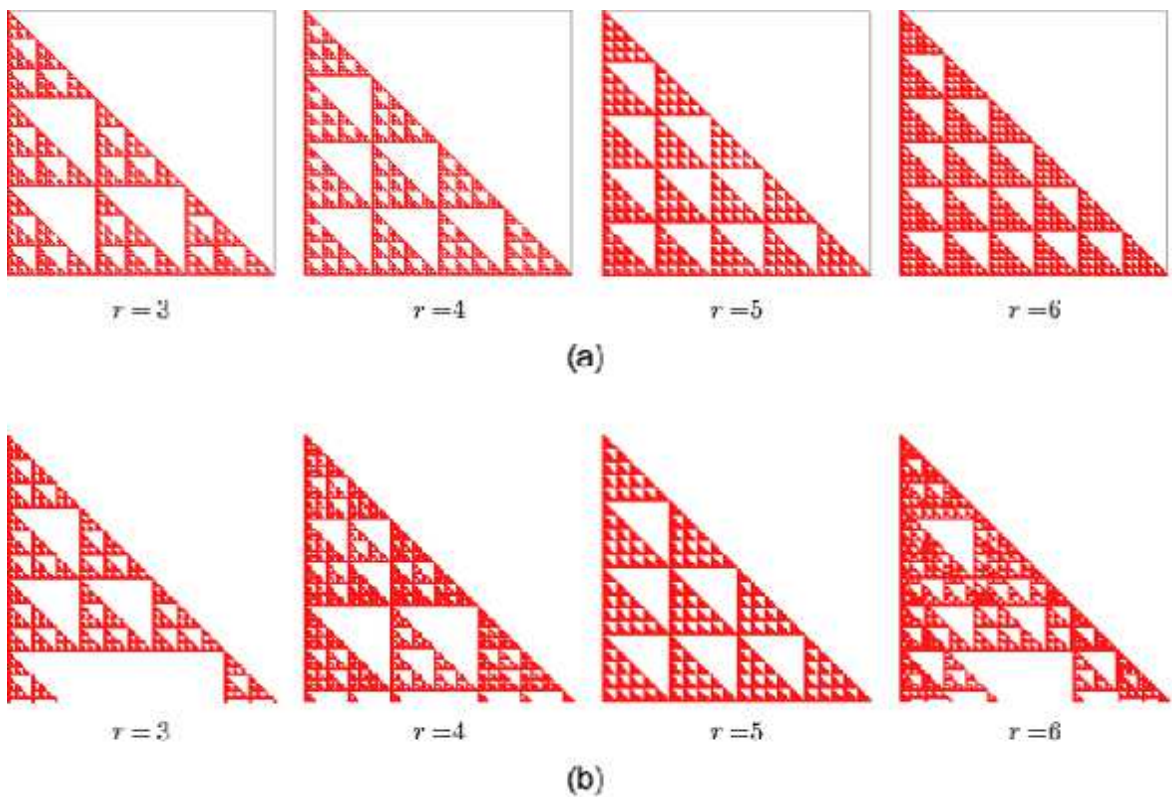}
\caption{
(a) The delay plots of the coins growing geometrically
(\denomi{1}, \denomi{r}, \denomi{r^2}, $\cdots$) for $r=3,4,5,$ and 6.
They all exhibit ordered shapes like the discrete Sierpinski gasket.
Since we compare only the shapes, the scales are omitted from the plots.
(b) The Pascal-Sierpinski gasket of modulo $r$
for the case $r=3,4,5$, and 6.
The upper part of 100 rows are shown.
Pascal-Sierpinski gaskets are similar to the delay plots for $r=3$ and 5,
and are different for $r=4$ and 6.
}
\label{fig:compare}
\end{figure}

In Fig. \ref{fig:compare} (a),
we show the delay plots corresponding to $r=3,4,5$, and 6.
The shapes of them are quite similar to the Sierpinski gasket.
In contrast, the Pascal-Sierpinski gaskets are more complicated
(Fig. \ref{fig:compare} (b)).
In the case of $r=3$ and 5,
the Pascal-Sierpinski gaskets are similar to the corresponding delay plots.
However, the Pascal-Sierpinski gaskets of modulo $r=4$ and 6
exhibit disordered structure,
and they looks different from the delay plots of such $r$.
According to the theory\cite{Frame} of the Pascal-Sierpinski gasket,
the Pascal-Sierpinski gasket possess the same structure
as the corresponding delay plot if $r$ is a prime number;
however, the pattern is more complicated if $r$ is a composite number.

Analogues of Fig. \ref{fig:iteration-binary}, Eq. \eqref{eq:dimension}, and
Eqs. \eqref{eq:area} and \eqref{eq:area-binary} have appeared
in some research articles of Pascal's triangle\cite{Gamelin, Peitgen, Peitgen2}.
Hence, we note that the minimal payment is a real-world counterpart
of such purely mathematical objects.

\section{Discussion}
\begin{figure}[p!] \centering
\includegraphics[width=130mm]{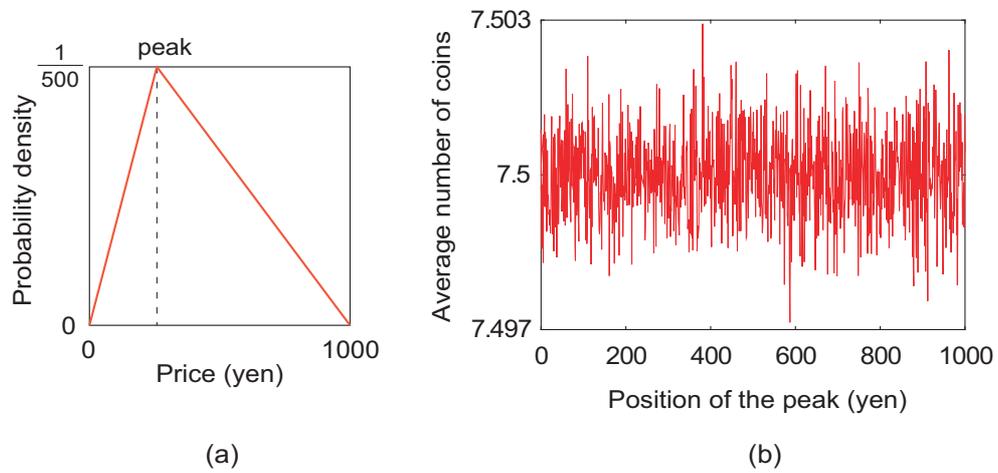}
\caption{
(a) The probability density function of a triangular random number.
The currency is the Japanese yen, so that
prices are distributed between 0 yen and 1000 yen.
The height $\frac{1}{500}$ is given by normalization.
(b) A numerical result of the average number of coins in a purse,
where the prices are generated by a triangular random number.
The average numbers are within $7.5\pm0.003$,
irrespective of the position of the peak of a triangular random number.
}
\label{fig:triangular}
\end{figure}

In the numerical simulations of the minimal payment in this paper,
we assume that the prices are distributed
randomly, independently, and uniformly.
Randomness and independentness are reasonable for real payment,
but uniformness is a debatable point.
In fact, the range of prices
depends on each person's life style or consumption behavior.
We simply model this non-uniform occurrence of prices
by using a triangular random number.
In this numerical calculation,
the Japanese yen (1000 yen bill; 500, 100, 50, 10, 5, and 1 yen coins)
was used as a currency,
and we iterated the minimal-payment processes $10^7+1$ times.
The prices are generated by a triangular random number
on the interval $[0, 1000]$,
and the position of the peak of the triangle is a control parameter
(Fig. \ref{fig:triangular} (a)).
The result is shown in Fig. \ref{fig:triangular} (b).
The average numbers of coins in the purse are within $7.5\pm0.003$,
which is irrespective of the position of the peak of triangle.
Thus, the average number of coins in a purse is still 7.5 pieces
also in this more realistic case.
We expect that
the average number of coins in a purse remains constant
over a wide class of probability distributions of prices.

The term `fractal time series' or `temporal fractal'
conventionally means that a time series has scale-invariance or
power-law spectrum\cite{Higuchi, Yamamoto, Caccia}.
On the other hand,
the time series of change in successive minimal payment
is another type of `fractal time series'
in the sense that its delay plot exhibits self-similarity.
We note that the power spectrum of a time series of change
does not exhibit power law.

The vector-like notation $(n_1,n_2,n_3)$ is a generalization
of the binary (or $r$-adic) expansion,
which is an essential tool in computer science\cite{Knuth, Baldoni}.
A cellular automaton was invented
as a prototype of digital computing.
Namely, our analysis originates from analogue payment,
and leads to digital fractals.
In the analysis, we assume that
a retailer gives back change with the least number of coins.
However,
real retailers may not give change with the least number of coins
because they may run out of coins of certain denominations,
or may feel that finding the least number is troublesome.
This causes the increase of the number of coins in a purse.
Thus, realization of the `minimal' requires the retailers' cooperation,
but
we emphasize that
we can avoid immoderate swelling of a purse at least
if we pay attention to our payment styles.

\section{Conclusion}
We demonstrated
that a payment method affects the number of coins left in a purse,
and showed how to minimize the number of coins in a purse after a payment.
We called this method the `minimal payment'.
By using the delay plot,
a time series of change under successive minimal-payment processes
a fine structure like the Sierpinski gasket.
The mechanism of this fine structure is reduced to a coupled inequalities
as in Eqs. \eqref{eq:area} and \eqref{eq:area-binary}.
We also compared delay plots
and the rule-60 cellular automaton and the Pascal-Sierpinski gasket.

\section*{Acknowledgments}
We are very greatful to Dr. Mitsugu Matsushita and Dr. Hiroto Kuninaka
for their beneficial comments.
One of the authors (K.Y.) is a Research Fellow
of the Japan Society for the Promotion of Science.

\clearpage
\newcommand{\refpaper}[6]{
	#1, #3, #4 (#6) #5.%
}
\newcommand{\refbook}[4]{
	#1, #2, #3, #4.%
}

\end{document}